\documentclass[aps,prd,reprint,superscriptaddress,nofootinbib,floatfix,longbibliography]{revtex4-2}
\usepackage[dvipsnames, table]{xcolor}
\usepackage{amssymb,amsmath,amsfonts,bm,slashed,soul,ragged2e,graphicx,epstopdf,hyperref,array}
\usepackage[utf8]{inputenc}
\usepackage{caption,subcaption}
\usepackage{multirow}
\enlargethispage{\baselineskip}
\hypersetup{colorlinks,linkcolor={blue},citecolor={blue},urlcolor={blue}}
\DeclareCaptionJustification{justified}{\justifying}
\everymath{\displaystyle}

\begin{document}

\title{Numerical parameterization of stationary axisymmetric black holes in a theory agnostic framework}

\author{Olzhas~Mukazhanov}
\email{olzhmuk@gmail.com}
\affiliation{Center for Astronomy and Astrophysics, Center for Field Theory and Particle Physics, and Department of Physics,\\
Fudan University, Shanghai 200438, China}

\author{Rittick~Roy}
\email{rittickrr@gmail.com}
\affiliation{Anton Pannekoek Institute for Astronomy, University of Amsterdam, 1098 XH, Amsterdam, The Netherlands}

\author{Temurbek Mirzaev}
\email{mtemurbek22@m.fudan.edu.cn}
\affiliation{Center for Astronomy and Astrophysics, Center for Field Theory and Particle Physics, and Department of Physics,\\
Fudan University, Shanghai 200438, China}

\author{Cosimo~Bambi}
\email{bambi@fudan.edu.cn}
\affiliation{Center for Astronomy and Astrophysics, Center for Field Theory and Particle Physics, and Department of Physics,\\
Fudan University, Shanghai 200438, China}
\affiliation{School of Natural Sciences and Humanities, New Uzbekistan University, Tashkent 100007, Uzbekistan}

\date{\today}

\begin{abstract}
\noindent 
The pursuit of a comprehensive theory of gravity has led to the exploration of various alternative models, necessitating a model-independent framework. The Konoplya-Rezzolla-Zhidenko (KRZ) parameterization offers a robust method for approximating stationary axisymmetric black hole spacetimes, characterized by a rapidly converging continued-fraction expansion. However, while analytical metrics benefit from this approach, numerical metrics derived from complex gravitational theories remain presenting computational challenges. Bridging this gap, we propose a method for a numerical KRZ parameterization, tested and demonstrated on pseudo-numerical Kerr and Kerr-Sen spacetimes. Our approach involves constructing numerical grids to represent metric coefficients and using the grids for fitting the parameters up to an arbitrary order. We analyze the accuracy of our method across different orders of approximation, considering deviations in the metric functions and shadow images. In both Kerr and Kerr-Sen cases, we observe rapid convergence of errors with increasing orders of continued fractions, albeit with variations influenced by spin and charge. Our results underscore the potential of the proposed algorithm for parameterizing numerical metrics, offering a pathway for further investigations across diverse gravity theories.
\end{abstract}

\maketitle

\section{Introduction}

Since its proposal over a century ago \cite{Einstein}, Einstein's theory of gravity has found applications across a multitude of astrophysical phenomena in our Universe. Through the years, it has solidified its position as the standard model for describing spacetime in the presence of gravitational fields. While predominantly successful in weak-field tests \cite{Will}, only recently have the strong-field predictions of Einstein's gravity, also known as the general theory of relativity (GR), become subject to various testable methods \cite{Bambi, Abbott19, EHT}. The proliferation of alternative gravity theories, which aim to address deficiencies of GR concerning observations such as dark matter and dark energy, or extend GR to resolve issues like the quantization of gravity and the curvature singularity, underscores the importance of scrutinizing GR's strong-field predictions using the latest techniques and technologies.

Black holes (BHs), found abundantly throughout our Universe, serve as ideal testing grounds for theories of gravity due to the intense gravitational fields surrounding them. In the framework of GR, under typical astrophysical conditions, BHs are characterized by a few key parameters, their mass and spin, rendering them Kerr black holes \cite{Kerr}. The Kerr hypothesis posits that astrophysical BHs adhere to the Kerr metric, an assumption that defines their simplicity in GR (for detailed conditions and assumptions, see Ref. \cite{Chrusciel}). Alternative gravity theories often introduce additional parameters, causing deviations from the Kerr solution.

Exploring the effects of BHs through observations has been a notable pursuit in physics, offering the potential for uncovering fascinating phenomena. Various methods employed include X-ray spectroscopy (leading to the first measurements of BH spin \cite{Dabrowski, Young}), gravitational wave interferometry (resulting in the first observation of BH coalescence \cite{Abbott16}), and BH imaging (providing the inaugural capture of an image near the BH horizon \cite{Akiyama_2019}).

In light of the vast variety of alternative gravity theories and the ongoing quest for a definitive theory, there is a compelling need to develop a model-independent framework. Several models have been suggested in the past \cite{Johannsen, Rezzolla, KRZ}. In particular, the Konoplya-Rezzolla-Zhidenko (KRZ) approach introduces a robust and generic parameterization to approximate stationary axisymmetric black hole spacetimes. Their methodology involves a rapidly converging continued-fraction expansion in a compactified radial coordinate allowing for approximations of high accuracy with only few parameters. The efficacy of this approach was demonstrated in \cite{Younsi} through a comparison of shadow images across various spacetimes.

However, complex gravitational theories that extend beyond General Relativity, lacking simple or exact solutions, often can only be described via numerical metrics that cannot be parameterized directly. Contrasting with analytical metrics, numerical metrics are crafted via computational simulations. By discretizing spacetime into a mesh or grid and applying numerical methods, one can approximate the values of the metric tensor across different points in spacetime. Although numerical metrics play an important role in exploring complex situations devoid of exact solutions, they fall short in computational efficiency when compared to their analytical counterparts, which demand less extensive resources. Therefore, a parameterization technique compatible with numerical metrics would combine the complexity of numerical spacetimes with the operational efficiency of analytical solutions.

In this paper, we present a way to perform the numerical KRZ parameterization. We test how the approach works on pseudo-numerical Kerr and Kerr-Sen spacetimes and study the accuracy for different orders of approximation.

This paper is structured as follows: Section~\ref{KRZ} reviews the Konoplya-Rezzolla-Zhidenko parameterization approach; Section~\ref{Numerical} discusses how the parameterization can be done numerically; Section~\ref{Results} illustrates the accuracy of Kerr and Kerr-Sen black hole shadows constructed with the numerically parameterized metrics as compared to the analytical metrics; Section~\ref{Conclusion} presents conclusions and discusses potential areas for improvement and applications of the work.

\section{Konoplya-Rezzolla-Zhidenko Parameterization}\label{KRZ}

We employ the Konoplya-Rezzolla-Zhidenko approach to parameterize stationary, axisymmetric black holes in a general framework. For the sake of completeness, we give a brief review of the parameterization scheme. For more details, we refer the reader to the original papers by Rezzolla and Zhidenko \cite{Rezzolla}, and Konoplya, Rezzolla, and Zhidenko \cite{KRZ}. The line element for an axisymmetric spacetime possesses a timelike and a spacelike Killing vector, permitting the selection of coordinates $t$ and $\phi$ aligned with these vectors. A general BH metric tensor with a normalized mass $M=1$ is given by
\begin{widetext}
\begin{equation}
    ds^2 = -\frac{f(\rho,\vartheta)-\omega^2(\rho,\vartheta)\sin^2\vartheta}{\kappa^2(\rho,\vartheta)} dt^2 - 2\omega(\rho,\vartheta) \rho\sin^2\vartheta dtd\phi +\kappa^2(\rho,\vartheta) \rho^2\sin^2\vartheta d\phi^2 + \sigma(\rho,\vartheta) \left(\frac{\beta^2(\rho,\vartheta)}{f(\rho,\vartheta)} d\rho^2 + \rho^2 d\vartheta^2\right), \label{eq:orig}
\end{equation}
\end{widetext}
where $f(\rho,\vartheta)$, $\beta(\rho,\vartheta)$, $\sigma(\rho,\vartheta)$, $\kappa(\rho,\vartheta)$, and $\omega(\rho,\vartheta)$ are dimensionless functions dependent solely on the coordinates $\rho$ and $\vartheta$.

By the principle of general covariance, we can always select a different pair of coordinates, $\rho$ and $\vartheta$, that would describe the same spacetime. To resolve the ambiguity and achieve a distinctive parameterization, we choose such coordinates $(r ,\theta)$, following \cite{KRZ}, that the line element takes on the following form:
\begin{widetext}
\begin{equation}
    ds^2 = -\frac{N^2(r,\theta) - W^2(r,\theta)\sin^2\theta}{K^2(r,\theta)}dt^2 - 2W(r,\theta)r\sin^2\theta dtd\phi + K^2(r,\theta)r^2\sin^2\theta d\phi^2 + \Sigma(r,\theta) \left(\frac{B^2(r,\theta)}{N^2(r,\theta)}dr^2 + r^2d\theta^2\right), \label{eq:new}
\end{equation}
\end{widetext}
where $N(r,\theta)$, $W(r,\theta)$, $K(r,\theta)$, $B(r,\theta)$ are arbitrary functions and
\begin{equation}
    \Sigma(r,\theta) \equiv r^2 + a^2\cos^2\theta,
\end{equation}
with $a$ being a spin parameter. The conversion between the coordinates is done by solving the set of equations:
\begin{subequations}\label{eq:main}
\begin{align}
    N^2(r,\theta)r^2\sin^2\theta &= f(\rho,\vartheta)\rho^2\sin^2\vartheta, \label{eq:sub1} \\
    W(r,\theta)r\sin^2\theta &= \omega(\rho,\vartheta)\rho\sin^2\vartheta, \label{eq:sub2} \\
    K^2(r,\theta)r^2\sin^2\theta &= \kappa^2(\rho,\vartheta)\rho^2\sin^2\vartheta, \label{eq:sub3} \\
    \frac{N^2(r,\theta)}{\Sigma(r,\theta)B^2(r,\theta)} &= \frac{1}{\sigma(\rho,\vartheta)} \left(\frac{f(\rho,\vartheta)}{\beta^2(\rho,\vartheta)} \frac{\partial r}{\partial\rho} \frac{\partial r}{\partial\rho} + \frac{1}{\rho^2} \frac{\partial r}{\partial\vartheta} \frac{\partial r}{\partial\vartheta}\right), \label{eq:sub4} \\
    \frac{1}{\Sigma(r,\theta)r^2} &= \frac{1}{\sigma(\rho,\vartheta)} \left(\frac{f(\rho,\vartheta)}{\beta^2(\rho,\vartheta)} \frac{\partial\theta}{\partial\rho} \frac{\partial\theta}{\partial\rho} + \frac{1}{\rho^2} \frac{\partial\theta}{\partial\vartheta} \frac{\partial\theta}{\partial\vartheta}\right), \label{eq:sub5} \\
    0 &= \frac{f(\rho,\vartheta)}{\beta^2(\rho,\vartheta)} \frac{\partial r}{\partial\rho} \frac{\partial\theta}{\partial\rho} + \frac{1}{\rho^2} \frac{\partial\theta}{\partial\vartheta} \frac{\partial r}{\partial\vartheta}. \label{eq:sub6}
\end{align}
\end{subequations}

For further convenience, we compactify the radial and angular coordinates:
\begin{equation}
    x \equiv 1 - \frac{r_0}{r}, \hspace{1cm}
    y \equiv \cos\theta,
\end{equation}
where $r_0$ denotes the horizon radius within the equatorial plane. Here, $x$ spans from $0$ at the horizon to $1$ at spatial infinity. Consequently, we express the metric functions in terms of $x$:
\begin{subequations}\label{eq:funcs}
\begin{align}
    \Sigma &= 1 + \frac{a^2}{r_0^2}(1-x)^2y^2, \\
    W &= \sum_{i=0}^\infty \frac{W_i(x)y^i}{\Sigma}, \\
    K^2 - \frac{aW}{r} &= 1 + \sum_{i=0}^\infty \frac{K_i(x)y^i}{\Sigma}, \\
    N^2 &= xA_0(x) + \sum_{i=1}^\infty A_i(x)y^i, \\
    B &= 1 + \sum_{i=0}^\infty B_i(x) y^i,
\end{align}
\end{subequations}
where
\begin{subequations}\label{eq:Wix}
\begin{align}
    W_i(x) &= w_{i0}(1-x)^2 + \tilde{W}_i(x)(1-x)^3, \\
    K_i(x) &= k_{i0}(1-x)^2 + \tilde{K}_i(x)(1-x)^3, \\
    A_0(x) &= 1 - \epsilon_0(1-x) + (a_{00} - \epsilon_0 + k_{00})(1-x)^2 \nonumber \\
        &\quad + \tilde{A}_0(x)(1-x)^3, \\
    A_{i>0}(x) &= K_i(x) + \epsilon_i(1-x)^2 + a_{i0}(1-x)^3 \nonumber \\
        &\quad + \tilde{A}_i(x)(1-x)^4, \\
    B_i(x) &= b_{i0}(1-x) + \tilde{B}_i(x)(1-x)^2.
\end{align}
\end{subequations}
The tilded functions $\tilde{W}_i(x)$, $\tilde{K}_i(x)$, $\tilde{A}_i(x)$, and $\tilde{B}_i(x)$ describe the black hole metric near its horizon via Pad\'{e} approximants which are given as:
\begin{subequations}\label{eq:Pade}
\begin{align}
    \tilde{W}_i(x) &= \dfrac{w_{i1}}{1 + \dfrac{w_{i2}x}{1+\dfrac{w_{i3}x}{1+...}}}, \\
    \tilde{K}_i(x) &= \dfrac{k_{i1}}{1 + \dfrac{k_{i2}x}{1+\dfrac{k_{i3}x}{1+...}}}, \\
    \tilde{A}_i(x) &= \dfrac{a_{i1}}{1 + \dfrac{a_{i2}x}{1+\dfrac{a_{i3}x}{1+...}}}, \\
    \tilde{B}_i(x) &= \dfrac{b_{i1}}{1 + \dfrac{b_{i2}x}{1+\dfrac{b_{i3}x}{1+...}}}.
\end{align}
\end{subequations}

\section{Numerical Parameterization}\label{Numerical}

The functions described in Eqs.~(\ref{eq:funcs}) and (\ref{eq:Pade}) are, in principle, expanded indefinitely. For numerical purposes, we truncate the $y$-expansion at power $m$ and the continued fractions at the $n$th term.

We normalize the black hole's mass to $1$ and construct numerical metrics as two-dimensional grids spanning $N_r$ points along the $r$ dimension and $N_\theta$ points along the $\theta$ dimension. Each grid point is associated with four metric coefficients: $g_{tt}$, $g_{t\phi}$, $g_{rr}$, and $g_{\phi\phi}$ expressed in the coordinates $(r, \theta)$ that ensure
\begin{equation}
    g_{\theta\theta} = r^2 + a^2\cos^2\theta,
\end{equation}
thereby removing the necessity for a coordinate transformation. Then, Eqs.~(\ref{eq:funcs}) can be rewritten in terms of metric tensor coefficients as:
{
\allowdisplaybreaks
\begin{subequations}\label{eq:numwix}
\begin{align}
    \sum_i^mW_i(x)y^i &= -\frac{g_{t\phi}}{r\sin^2\theta}\Sigma, \\
    \sum_i^mK_i(x)y^i &= \left(\frac{g_{\phi\phi}+ag_{t\phi}}{r^2\sin^2\theta} - 1\right)\Sigma, \\
    xA_0(x) + \sum_{i=1}^m A_i(x) y^i &= \frac{g_{t\phi}^2 - g_{tt}g_{\phi\phi}}{r^2\sin^2\theta}, \\
    \sum_{i=0}^m B_i(x)y^i &= \frac{1}{r\sin\theta}\sqrt\frac{g_{rr}(g_{t\phi}^2-g_{tt}g_{\phi\phi})}{\Sigma} - 1,
\end{align}
\end{subequations}
}
The value of $m$ corresponds to the highest power of $\cos\theta$ in the expansion, with higher orders being truncated. Additionally, assuming the spacetime's reflection symmetry across the equatorial plane, we consider only even powers of $y$ in the fit.

\begin{figure}
    \centering
    \includegraphics[width=0.45\textwidth]{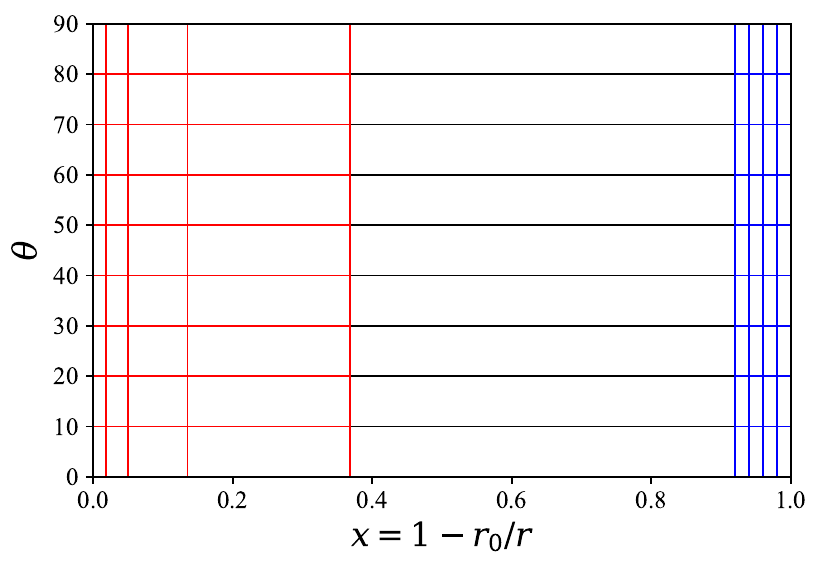}
    \caption{Numerical grid with two dimensions: compactified radial coordinate $x$ (unitless) and angular $\theta$ (degrees). The red points are used to fit the Pad\'{e} approximants near the horizon. The blue points are used to determine the asymptotic parameters.}
    \label{fig:grid}
\end{figure}

The fitting parameters are divided into two groups: asymptotic parameters and strong-field (Pad\'{e}) parameters. The former category includes parameters $w_{i0}$, $k_{i0}$, $a_{i0}$, $b_{i0}$, and $\epsilon_i$, i.e., all parameters that are outside the continued fractions. Consequently, the strong-field parameters include all the Pad\'{e} terms. This division necessitates the construction of a numerical grid that effectively spans both distant and proximal regions with respect to the black hole. An illustrative example of such a grid is presented in Figure \ref{fig:grid}.

The initial phase in the fitting process involves identifying the values of $W_i(x_j)$, $K_i(x_j)$, $A_i(x_j)$, and $B_i(x_j)$, where $j$ ranges across all radial points $N_r$ on the grid with dimensions $(N_r,N_\theta)$. This identification involves performing a polynomial fit in the angular direction, using $y^2=\cos^2\theta$ as the polynomial variable. Each radial coordinate $x_j$ is associated with $N_\theta$ points on the grid, which are used in the fitting process. Increasing the value of $N_\theta$ can reduce the error in computing $W_i(x_j)$, $K_i(x_j)$, $A_i(x_j)$, and $B_i(x_j)$.

After computing $W_i(x_j)$, $K_i(x_j)$, $A_i(x_j)$, and $B_i(x_j)$ at all radii of the grid, the next step is determining asymptotic and Pad\'{e} parameters. As seen from Eqs.~(\ref{eq:Wix}), the vicinity of the black hole features both groups of parameters, while the far region depends predominantly on the asymptotic ones. Therefore, determining the asymptotic parameters in the far region first proves to be more practical. At substantial radii ($x\to1$), the continued fractions effectively become constants and can be approximated as:
{
\allowdisplaybreaks
\begin{subequations}
\begin{align}
    \tilde{W}_i(x\to1) &\approx \dfrac{w_{i1}}{1 + \dfrac{w_{i2}}{1+\dfrac{w_{i3}}{1+...}}}, \\
    \tilde{K}_i(x\to1) &\approx \dfrac{k_{i1}}{1 + \dfrac{k_{i2}}{1+\dfrac{k_{i3}}{1+...}}}, \\
    \tilde{A}_i(x\to1) &\approx \dfrac{a_{i1}}{1 + \dfrac{a_{i2}}{1+\dfrac{a_{i3}}{1+...}}}, \\
    \tilde{B}_i(x\to1) &\approx \dfrac{b_{i1}}{1 + \dfrac{b_{i2}}{1+\dfrac{b_{i3}}{1+...}}}.
\end{align}
\end{subequations}
}
Subsequently, Eqs. (\ref{eq:Wix}) can be approximated as ordinary polynomials of $1-x$:
\begin{subequations}
\begin{align}
    W_i(x\to1) &\approx w_{i0}(1-x)^2 + \tilde{W}_i(1)(1-x)^3, \\
    K_i(x\to1) &\approx k_{i0}(1-x)^2 + \tilde{K}_i(1)(1-x)^3, \\
    A_0(x\to1) &\approx 1 - \epsilon_0(1-x) \nonumber \\
        &\quad + (a_{00} - \epsilon_0 + k_{00})(1-x)^2 \\
        &\quad + \tilde{A}_0(1)(1-x)^3, \nonumber \\
    A_{i>0}(x\to1) &\approx K_i(x) + \epsilon_i(1-x)^2 + a_{i0}(1-x)^3 \nonumber \\
        &\quad + \tilde{A}_i(1)(1-x)^4, \\
    B_i(x\to1) &\approx b_{i0}(1-x) + \tilde{B}_i(1)(1-x)^2.
\end{align}
\end{subequations}
Since these functions are, effectively, polynomials in terms of $1-x$, it is now possible to perform a polynomial fit in the same manner as it was done in the angular direction. The previously computed values of $W_i(x_j)$, $K_i(x_j)$, $A_i(x_j)$, and $B_i(x_j)$ are used as fitting data points, where $j$ spans only the large radii in the asymptotic region of the grid, as visually represented by the blue region in Figure~\ref{fig:grid}. The fitting process yields numerical estimates for the parameters $w_{i0}$, $k_{i0}$, $a_{i0}$, $\epsilon_i$, and $b_{i0}$.

From Eqs.~(\ref{eq:Wix}), combined with the knowledge of the asymptotic parameters, we can deduce the values of the tilded functions $\tilde{W}_i(x_j)$, $\tilde{K}_i(x_j)$, $\tilde{A}_i(x_j)$, and $\tilde{B}_i(x_j)$ in the black hole's vicinity, where $j$ ranges across small radii depicted in red on Figure~\ref{fig:grid}.
\begin{subequations}\label{eq:Wixnear}
\begin{align}
    \tilde{W}_i(x)(1-x)^3 &= W_i(x) - w_{i0}(1-x)^2, \\
    \tilde{K}_i(x)(1-x)^3 &= K_i(x) - k_{i0}(1-x)^2, \\
    \tilde{A}_0(x)(1-x)^3 &= A_0(x) - 1 + \epsilon_0(1-x) \nonumber \\
        &\quad - (a-{00}-\epsilon_0+k_{00})(1-x)^2, \\
    \tilde{A}_{i>0}(x)(1-x)^4 &= A_i(x) - K_i(x) - \epsilon_i(1-x)^2 \nonumber \\
        &\quad - a_{i0}(1-x)^3, \\
    \tilde{B}_i(x)(1-x)^2 &= B_i(x) - b_{i0}(1-x).
\end{align}
\end{subequations}
From Eqs.~(\ref{eq:Pade}) and (\ref{eq:Wixnear}), we determine the Pad\'{e} parameters. The tilded functions are not polynomials in their explicit form. Nevertheless, it is possible to rearrange the continued fractions into a polynomial form, enabling the construction of a set of linear equations, which can then be subjected to a polynomial fitting procedure. The complexity of the equations significantly increases with the Pad\'{e} order $n$. Alternatively, in this paper we utilized numerical solving algorithms to determine the strong-field parameters.

\section{Results}\label{Results}

In this work, the Kerr and Kerr-Sen metric values are expressed on a grid and are constructed as pseudo-numerical metrics for testing purposes. We compare the levels of accuracy for different orders $n$ of the continued fractions in both spacetimes. In all tests, we fix the highest cosine term in the polar expansion as $\cos^2\theta$, since both metrics do not require higher-order cosine terms. Additionally, we are not interested in analyzing the parameterization accuracy in the polar direction for different orders $m$, since the polar parameterization is rather trivial and does not require an in-depth analysis. Thus, we only consider $m=2$ in our tests.

After computing the parameters, we construct parameterized metric coefficients $g_{\mu\nu}^\text{n}$ and compare them to the analytical coefficients $g_{\mu\nu}^\text{a}$. The relative error is computed as
\begin{equation}
    \varepsilon_{\mu\nu}^\text{metric} = \left|1 - \frac{g_{\mu\nu}^\text{n}}{g_{\mu\nu}^\text{a}}\right|.
\end{equation}
Apart from the metric values, we compare shadows which are constructed by "backlit" photons revolving multiple times around the black hole before reaching the observer screen. The screen is divided into 360 degrees and the shadow error is computed as:
\begin{equation}
    \varepsilon_\text{sh} = \sqrt{\frac{\sum_i (r_i^\text{n} - r_i^\text{a})^2}{\sum_i \left(r_i^\text{a}\right)^2}},
\end{equation}
where $r_i^\text{n}$ and $r_i^\text{a}$ are numerical and analytical shadow radii on the screen at every degree.

\subsection{Kerr Parameterization}

The Kerr spacetime serves as a foundational example where the metric is precisely parameterized using a minimal set of parameters \cite{KRZ}. The line element in Boyer-Lindquist (BL) coordinates requires parameters only up to $\cos^2\theta$ ($m=2$) and up to the third 
Pad\'{e} order ($n=3$). Additionally, the BL coordinates coincide with the coordinates of choice in the KRZ parameterization, since
\begin{equation}
    g_{\theta\theta}^\text{Kerr} = r^2 + a^2\cos^2\theta.
\end{equation}
Therefore, no transformations are necessary for the construction of the numerical grid.

For testing purposes, following the procedure detailed in Section~\ref{Numerical}, we numerically compute continued fractions with orders $n=1,2,3$. We do not consider parameters corresponding to terms higher than $\cos^2\theta$ and $n=3$, since higher orders do not provide non-vanishing values.

\begin{figure}[t]
    \centering
    \includegraphics[width=0.45\textwidth]{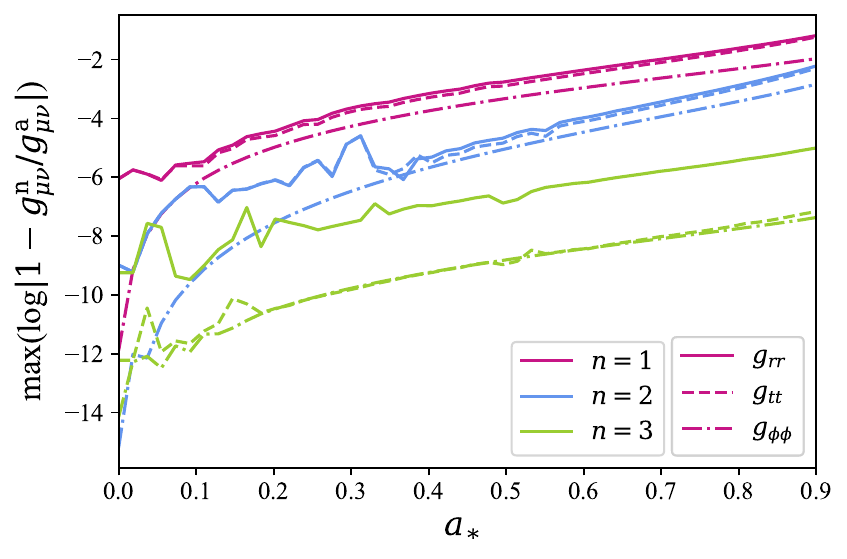}
    \caption{Maximum relative error of the numerically parameterized Kerr metric functions versus spin $a_*$ for different orders of continued fractions $n$. The order of polar expansion is $m=2$. The maximum is taken over the range from $r_\text{min}=1.4r_0$ to infinity. The minimum radius $r_\text{min}$ is taken in such a way that avoids $g_{tt}=0$ while still including shadow and ISCO radii. The errors for $g_{t\phi}$ and $g_{\theta\theta}$ are universally zero and not included in the plot.}
    \label{fig:Kerr_metric_error}
\end{figure}

After computing the parameters, we construct parameterized metric coefficients $g_{\mu\nu}^\text{n}$ and compare them to the analytical coefficients $g_{\mu\nu}^\text{a}$. Regions closer to the horizon are associated with higher deviations. Additionally, the relative error blows up at the ergosphere where $g_{tt}=0$. To avoid such numerical singularities, we consider the error over the region $x \in (0.3, 1)$, where $x=0.3$ corresponds to $r\approx1.4r_0$. This radius excludes the ergosphere but includes the shadow and ISCO radii for all the considered spins. Figure~\ref{fig:Kerr_metric_error} depicts the maximum relative error in the region for the metric functions $g_{rr}$, $g_{tt}$, and $g_{\phi\phi}$. The error for $g_{t\phi}$ is universally zero since the Kerr metric does not require any continued fractions to describe this function. The error for $g_{\theta\theta}$ is zero by construction, as we define it to be $g_{\theta\theta} = r^2+a^2\cos^2\theta$.

The upward trend in error versus spin is evident from Figure~\ref{fig:Kerr_metric_error}. The reason is as follows: as the spin increases, the horizon radius $r_0$ shrinks and becomes exposed to stronger gravitational fields, rendering numerical approximations less accurate. Additionally, since the case $n=3$ is supposed to describe the Kerr spacetime exactly, its error trend illustrated on the figure is contributed by numerical errors of the algorithm that are not associated with the KRZ parameterization technique. Additionally, these small errors may occasionally manifest as spikes below $10^{-4}$, which are visible on the plot. The maximum errors at $a_*=0.9$ for $n=1$ and $n=2$ remain below 10\% and 1\%, respectively.

Table~\ref{tab:Kerr_param_err} displays the accuracy of all numerically computed parameters that do not vanish in the Kerr spacetime with spin $a_*=0.998$. The error is computed relative to the analytical values. Asymptotic coefficients are generally independent of the Pad\'{e} order $n$, except for $a_{20}$, which approaches the expected value with higher orders. Similarly, strong-field parameters become more accurate at higher $n$.

\begin{table}[t]
    \centering
    \begin{tabular}{|c|c|c|c|}
    \hline
    \multicolumn{4}{|c|}{$\mathbf{a_* = 0.998}$} \\
    \hline
    \multirow{2}{*}{\textbf{Parameter}} & \multicolumn{3}{c|}{\textbf{Error (\%)}} \\
    \cline{2-4}
     & $n=1$ & $n=2$ & $n=3$ \\
    \hline \rowcolor[RGB]{204,229,255}
    $w_{00}$ & $1\times10^{-7}$ & $1\times10^{-7}$ & $1\times10^{-7}$ \\
    \hline \rowcolor[RGB]{204,229,255}
    $k_{00}$ & $5\times10^{-8}$ & $5\times10^{-8}$ & $5\times10^{-8}$ \\
    $k_{21}$ & $0.9$ & $7\times10^{-2}$ & $9\times10^{-4}$ \\
    $k_{22}$ & & $9$ & $1\times10^{-4}$ \\
    $k_{23}$ & & & $1\times10^{-3}$ \\
    \hline \rowcolor[RGB]{204,229,255}
    $\epsilon_0$ & $6\times10^{-8}$ & $5\times10^{-8}$ & $5\times10^{-8}$ \\
    \rowcolor[RGB]{204,229,255}
    $a_{20}$ & $12$ & $0.4$ & $6\times10^{-2}$ \\
    $a_{21}$ & $26$ & $0.9$ & $0.1$ \\
    \hline
    \end{tabular}
    \caption{Accuracy of numerically computed parameters that are non-vanishing in the Kerr spacetime. All the parameters not included in the table are zero. The coefficients highlighted in blue correspond to asymptotic parameters and are independent of the Pad\'{e} order $n$. Some cells are empty because lower orders do not support parameters of higher orders.}
    \label{tab:Kerr_param_err}
\end{table}

\begin{figure}[t]
    \centering
    \includegraphics[width=0.45\textwidth]{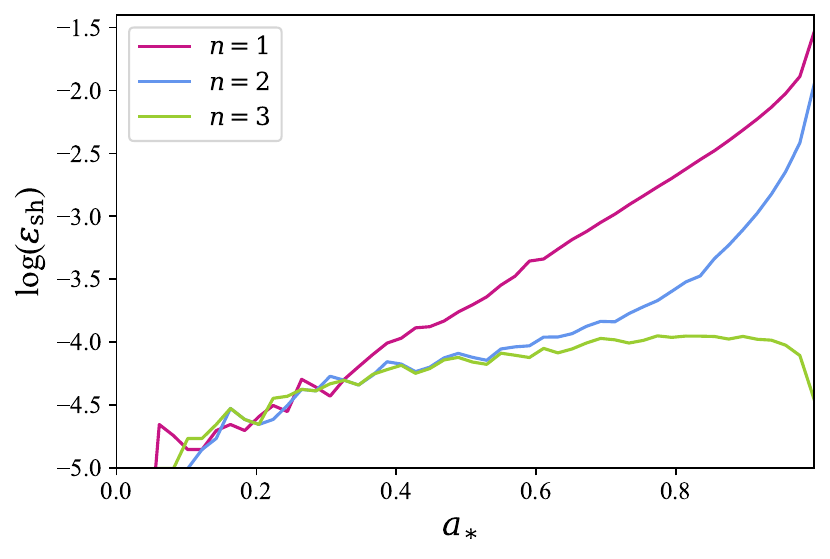}
    \caption{Logarithm of Kerr shadow error versus spin $a_*$ for different orders of continued fractions $n$. The order of polar expansion is $m=2$. The inclination angle is $\iota=30^\circ$.}
    \label{fig:Kerr_inc}
\end{figure}

The shadow error behavior versus spin $a_*$ is depicted on Figure~\ref{fig:Kerr_inc}. As expected, the trend is upward and deviations significantly decrease with every subsequent order of continued fractions. In particular, at $n=3$, the error accumulates due to numerical imprecision in the parameter calculations and the raytracing routines.

\begin{figure}[t]
    \centering
    \includegraphics[width=0.45\textwidth]{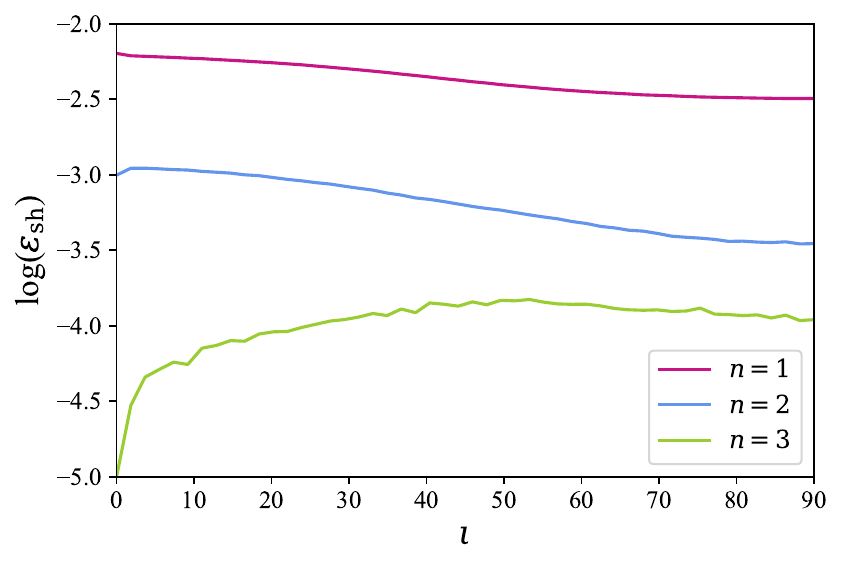}
    \caption{Logarithm of Kerr shadow error versus inclination angle $\iota$ for different orders of continued fractions $n$. The order of polar expansion is $m=2$. The spin is $a_*=0.9$.}
    \label{fig:Kerr_spin}
\end{figure}

Figure~\ref{fig:Kerr_spin} illustrates how the error depends on the inclination angle $\iota$. In the equatorial plane, where $\cos\theta=0$, the deviations are minimal. In the Kerr case specifically, the equatorial plane can be parameterized without the need for continued fractions, resulting in the exact ISCO radius for all values of $n$.

\subsection{Kerr-Sen Parameterization}

Sen's work \cite{Sen} introduced a solution for a rotating charged black hole by adapting the Kerr solution within the context of string theory's action in four dimensions. This adaptation leverages the intrinsic characteristics of string theory to extend the classical Kerr solution to include electrical charge, thus leading to the formulation of the Kerr-Sen (KS) metric.

\begin{figure}
    \centering
    \includegraphics[width=0.45\textwidth]{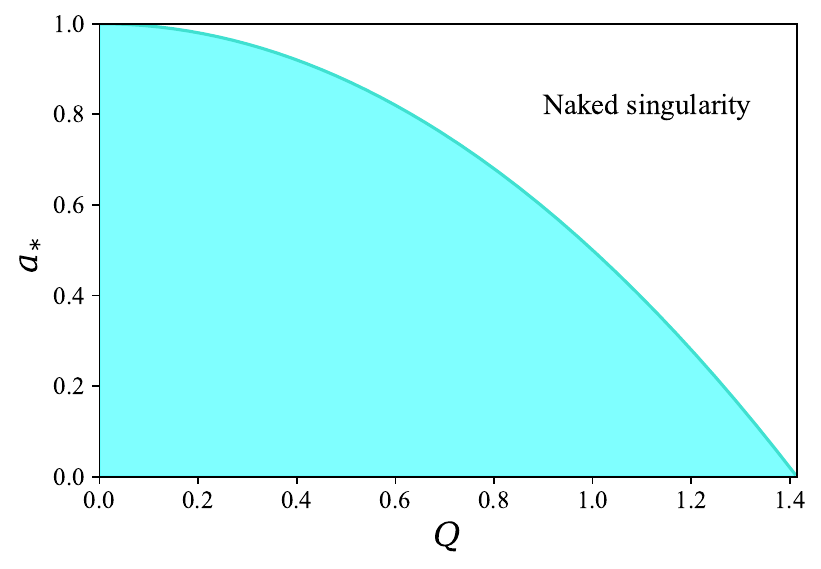}
    \caption{The shaded region includes all configurations of spin and charge that correspond to a black hole solution. In the unshaded region, the spacetime admits a naked singularity.}
    \label{fig:KS_sin}
\end{figure}

In the Boyer-Lindquist-like coordinates, the term $g_{\theta\theta}^\text{KS}$ is expressed as:
\begin{equation}
    g_{\theta\theta}^\text{KS} = \rho(\rho+Q^2) + a^2\cos^2\theta,
\end{equation}
where $Q$ is the dilaton charge and $a$ is the spin. Since $g_{\theta\theta}^\text{KS}$ does not follow the conventions outlined in Section~\ref{KRZ}, a transformation of the radial coordinate is required. Specifically, we modify the radial coordinate as follows:
\begin{subequations}
\begin{align}
    r &= \frac{Q^2}{2} \sqrt{\left(1+\frac{2\rho}{Q^2}\right)^2-1}, \\
    \rho &= \frac{Q^2}{2}\left(\sqrt{1+\frac{4r^2}{Q^4}} - 1\right).
\end{align}
\end{subequations}
The event horizon radius is expressed as:
\begin{equation}
    r_0 = \sqrt{\left(1 + \sqrt{\left(1-\frac{Q^2}{2}\right)^2 - a^2}\right)^2 - \frac{Q^4}{4}}.
\end{equation}
This relation highlights how the horizon is influenced by both the black hole's spin and charge. Figure~\ref{fig:KS_sin} graphically delineates the configurations of spin and charge that admit either black hole or naked singularity solutions in the theory.

The Kerr-Sen black hole cannot be parameterized exactly with a finite number of parameters \cite{KRZ}. However, a truncated form can approximate the spacetime with an arbitrary level of accuracy that converges fast with subsequent orders of Pad\'{e} approximants.

On the other hand, similarly to the Kerr scenario, the Kerr-Sen metric is exactly expanded in the polar direction up to the second cosine order, $m=2$.

\begin{figure}[t]
    \centering
    \includegraphics[width=0.45\textwidth]{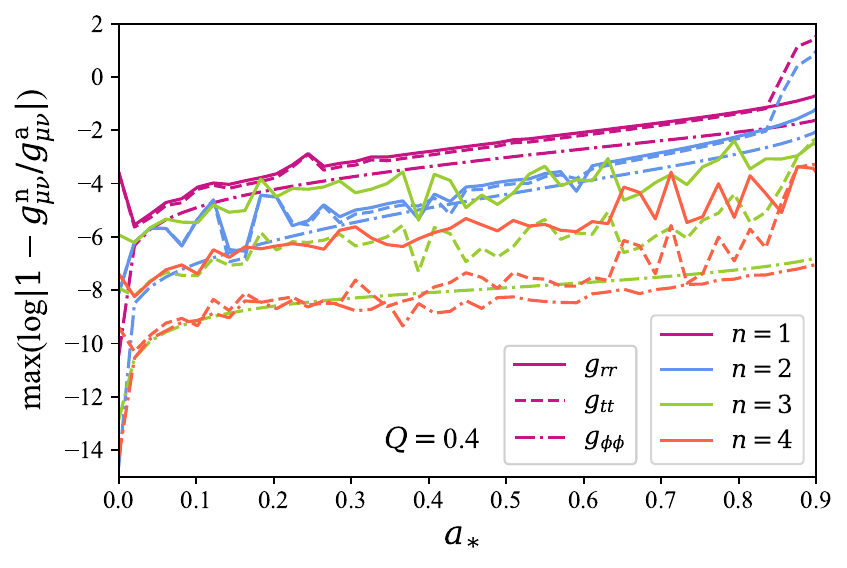}
    \caption{Maximum relative error of the numerically parameterized Kerr-Sen metric functions versus spin $a_*$ for different orders of continued fractions $n$. The order of polar expansion is $m=2$. The maximum is taken over the range from $r_\text{min}=1.4r_0$ to infinity. The minimum radius $r_\text{min}$ is taken in such a way that avoids $g_{tt}=0$ while still including shadow and ISCO radii. The error for $g_{\theta\theta}$ is universally zero and not included in the plot. The charge is $Q=0.4$.}
    \label{fig:KS_metric_0.4}
\end{figure}

\begin{figure}[t]
    \centering
    \includegraphics[width=0.45\textwidth]{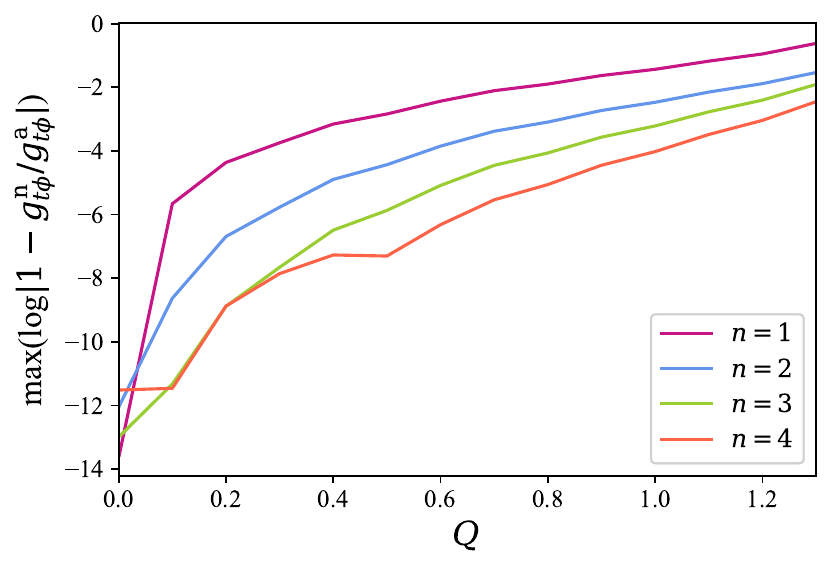}
    \caption{Maximum relative error of the numerically parameterized Kerr-Sen metric function $g_{t\phi}^\text{n}$ compared to the analytical counterpart $g_{t\phi}^\text{a}$ versus charge $Q$ for different orders of continued fractions $n$. The order of polar expansion is $m=2$. The maximum is taken over the range of all spins.}
    \label{fig:KS_gtp}
\end{figure}

In our tests, we parameterize the Kerr-Sen spacetime up to the second order of $\cos\theta$ ($m=2$) and consider four different orders of continued fractions, $n=1,2,3,4$. Analogously to the Kerr case, we observe the maximum deviations in the parameterized metric coefficients over the range $x \in (0.3, 1)$, where $x=0$ corresponds to $r\approx1.4r_0$. Figure~\ref{fig:KS_metric_0.4} depicts the errors for the metric coefficients $g_{rr}, g_{tt}$, and $g_{\phi\phi}$. We observe the spiky nature for errors below $10^{-4}$ resulting from numerical errors of the algorithm. As spin approaches extreme values, the errors start to diverge. This behavior is not observed in the Kerr case depicted in Figure~\ref{fig:Kerr_metric_error}, because the horizon remains relatively distant from the singularity. Figure~\ref{fig:KS_sin} shows that the configuration with $a=0.9$ and $Q=0.4$ is much closer to the naked singularity than the Kerr BH with the same spin. Additionally, $g_{t\phi}$ is no longer trivial in the presence of a charge $Q$. Figure~\ref{fig:KS_gtp} illustrates how accurate $g_{t\phi}$ is for different charges.

\begin{figure}[b]
    \centering
    \begin{subfigure}{\linewidth}
        \centering
        \includegraphics[width=0.9\linewidth]{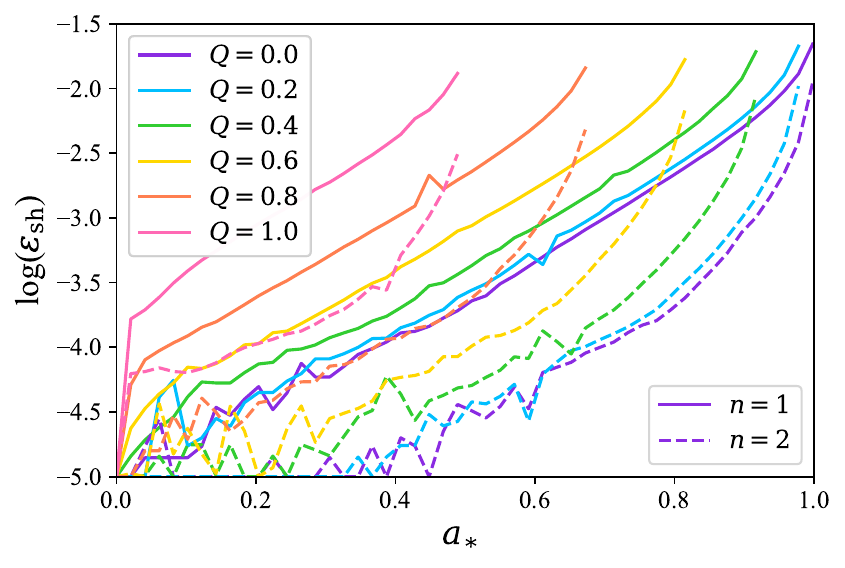}
        \caption{\centering $n=1,2$}
        \label{fig:KS_err_12}
    \end{subfigure}
    \begin{subfigure}{\linewidth}
        \centering
        \includegraphics[width=0.9\linewidth]{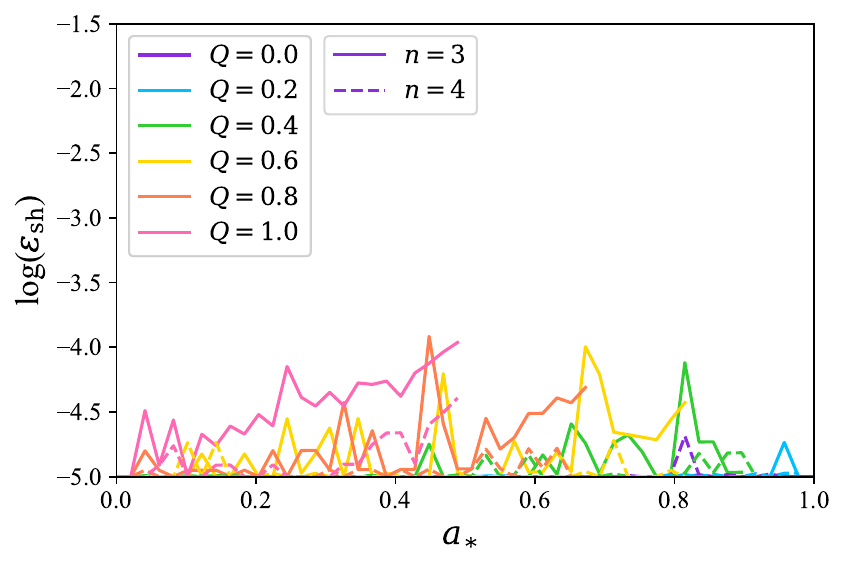}
        \caption{\centering $n=3,4$}
    \label{fig:KS_err_34}
    \end{subfigure}
    \caption{Kerr-Sen shadow error versus spin $a_*$ for different orders of continued fractions, and different charges $Q$. The order of polar expansion is $m=2$. The inclination angle is $\iota=30^\circ$. The trend is clear for $n=1$ and $n=2$. In the case of $n=3$ and $n=4$, errors reach the limit of numerical precision, where no clear trend is observed.}
    \label{fig:KS_err_1234}
\end{figure}

Shadow accuracy is shown on Figure~\ref{fig:KS_err_1234}. The sub-figure~\ref{fig:KS_err_12} illustrates a clear trend, which is also observed in the Kerr case. On the other hand, the sub-figure~\ref{fig:KS_err_34} shows that parameterization with $n=3$ and $n=4$ produces similar results, where we reach the limit of numerical accuracy and obtain curves with no particular trend.

\begin{figure}[t]
    \centering
    \includegraphics[width=0.45\textwidth]{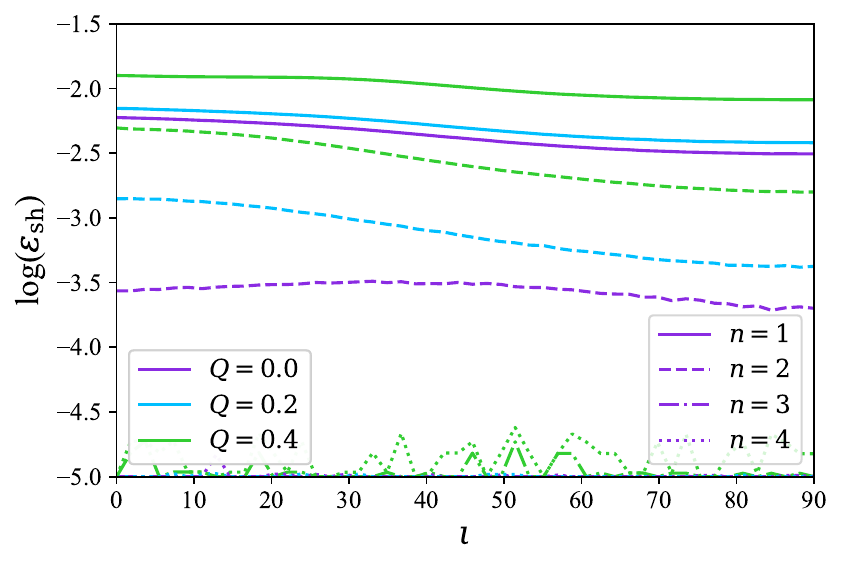}
    \caption{Kerr-Sen shadow error versus inclination angle $\iota$ for different orders of continued fractions $n$ and different charges $Q$. The order of polar expansion is $m=2$. The spin is $a_*=0.9$.}
    \label{fig:KS_err_spin}
\end{figure}

The errors remain consistent across the entire range of the inclination angle $\iota$, as depicted on Figure~\ref{fig:KS_err_spin}. Higher charges at a fixed spin correspond to a smaller horizon radius, yielding greater deviations from the expected images. At the order $n=3$ and below, the errors decrease to the limit of numerical precision, where no clear trend exists. The improvement at $n=4$ is marginal.

Table~\ref{tab:KS_param_err} illustrates the errors of all non-vanishing numerically computed parameters in the Kerr-Sen spacetime with a spin of $a_* = 0.4$ and charge $Q=1.0$. The selected black hole parameters produce an extreme scenario characterized by significant spin and charge. The error is evaluated relative to analytical values obtained through a direct expansion of the analytical metric. Similar to the Kerr case, asymptotic coefficients remain precise for all $n$. Generally, Pad\'{e} parameters exhibit increased accuracy with higher $n$, with the exception of $a_{2i}$, whose accuracy remains relatively stable as the Pad\'{e} order increases. This can be attributed to their lesser contribution to the overall spacetime.

\begin{table}[t]
    \centering
    \begin{tabular}{|c|c|c|c|c|}
    \hline
    \multicolumn{5}{|c|}{$\mathbf{a_* = 0.4 \quad Q = 1.0}$} \\
    \hline
    \multirow{2}{*}{\textbf{Parameter}} & \multicolumn{4}{c|}{\textbf{Error (\%)}} \\
    \cline{2-5}
     & $n=1$ & $n=2$ & $n=3$ & $n=4$ \\
    \hline \rowcolor[RGB]{204,229,255}
    $w_{00}$ & $5\times10^{-6}$ & $1\times10^{-5}$ & $5\times10^{-6}$ & $5\times10^{-6}$ \\
    $w_{01}$ & $0.2$ & $3\times10^{-3}$ & $2\times10^{-4}$ & $2\times10^{-4}$ \\
    $w_{02}$ & & $1$ & $3\times10^{-3}$ & $3\times10^{-5}$ \\
    $w_{03}$ & & & $2$ & $5\times10^{-2}$ \\
    $w_{04}$ & & & & 3 \\
    \hline \rowcolor[RGB]{204,229,255}
    $k_{00}$ & $5\times10^{-4}$ & $3\times10^{-4}$ & $3\times10^{-4}$ & $3\times10^{-4}$ \\
    $k_{21}$ & $0.6$ & $2\times10^{-2}$ & $5\times10^{-4}$ & $5\times10^{-4}$ \\
    $k_{22}$ & & $4$ & $8\times10^{-4}$ & $9\times10^{-4}$ \\
    $k_{23}$ & & & $0.2$ & $2\times10^{-2}$ \\
    $k_{24}$ & & & & $9$ \\
    \hline \rowcolor[RGB]{204,229,255}
    $\epsilon_0$ & $3\times10^{-5}$ & $3\times10^{-5}$ & $3\times10^{-5}$ & $3\times10^{-5}$ \\
    \rowcolor[RGB]{204,229,255}
    $a_{00}$ & $4\times10^{-5}$ & $3\times10^{-5}$ & $4\times10^{-5}$ & $4\times10^{-5}$ \\
    $a_{01}$ & $0.3$ & $3\times10^{-2}$ & $2\times10^{-5}$ & $2\times10^{-4}$ \\
    $a_{02}$ & & $8$ & $1\times10^{-2}$ & $4\times10^{-4}$ \\
    $a_{03}$ & & & $1$ & $2\times10^{-2}$ \\
    $a_{04}$ & & & & $1$ \\
    \hline \rowcolor[RGB]{204,229,255}
    $a_{20}$ & $5\times10^{-2}$ & $9\times10^{-3}$ & $2\times10^{-2}$ & $6\times10^{-2}$ \\
    $a_{21}$ & $8\times10^{-2}$ & $2\times10^{-2}$ & $4\times10^{-2}$ & $0.2$ \\
    $a_{22}$ & & $0.8$ & $0.2$ & $0.7$ \\
    $a_{23}$ & & & $5$ & $8$ \\
    $a_{24}$ & & & & $9$ \\
    \hline
    \end{tabular}
    \caption{Accuracy of numerically computed parameters that are non-vanishing in the Kerr-Sen spacetime. All the parameters not included in the table are zero. The coefficients highlighted in blue correspond to asymptotic parameters and are independent of the Pad\'{e} order $n$. Some cells are empty because lower orders do not support parameters of higher orders.}
    \label{tab:KS_param_err}
\end{table}

\section{Conclusion}\label{Conclusion}

We have developed an algorithm to perform the Konoplya-Rezzolla-Zhidenko parameterization on numerical stationary axisymmetric metrics and tested the technique on pseudo-numerical Kerr and Kerr-Sen spacetimes. The test metric is constructed as a two-dimensional grid, where the radial and polar dimensions range from the horizon to the spatial infinity and from the pole to the equatorial plane, respectively. We analyzed numerical deviations from the analytical metric functions and compared shadow images.

The errors of metric functions and shadows rapidly converge as the order of Pad\'{e} approximants increases. The strong-field parameterization parameters tend to approach their expected values as the Pad\'{e} order $n$ increases, while the asymptotic parameters remain accurate for all cases. As the horizon shrinks due to higher spin and/or higher charge, it becomes exposed to stronger gravitational curvature, necessitating a greater number of parameters for an accurate expansion at the horizon. However, even the most extreme cases require a small number of terms in the continued fractions.

An important point that is beyond the scope of this paper but is worth to be mentioned here nonetheless is the precision of the numerical data and how it might affect the parameterization accuracy. A numerical metric is expected to contain some inherent errors, which might also result in a non-spherical horizon. Since we employ polynomial fitting in the algorithm described in Section~\ref{Numerical}, an increase in grid points might help reduce some noise in the data, provided that the noise is unbiased. Similarly, if the horizon radius contains evenly spread deviations at different angles, averaging across multiple angular points will improve the radius precision.

The numerical 2D grid used in this paper describes the metrics in the conventional coordinates, $(r, \theta)$, that ensure $g_{\theta\theta} = r^2 + a^2\cos^2\theta$. In principle, a numerical metric can be represented as a 3D grid in arbitrary coordinates, potentially containing removable $\phi$-dependence coming from gauge effects. Therefore, some pre-processing needs to be done to convert such a metric into a 2D grid in the conventional coordinates. In future work, enhancing the algorithm's flexibility by incorporating an option for numerical coordinate transformations would streamline the process of metric preparation and enable analysis in alternative coordinate systems.

Apart from working solely with pseudo-numerical Kerr and Kerr-Sen spacetimes, it is important to conduct tests on a broader range of spacetimes derived from various gravity theories, such as EDGB \cite{EDGB, EDGBerr}, Horndeski/Galileon \cite{Horndeski, Galileon12, Galileon14}, or Nonlinear Electrodynamics theories \cite{Bronnikov, Ghosh}.

Additionally, beyond evaluating the parameterization's accuracy for black hole shadows, there is potential to broaden the scope of our tests to encompass other observational phenomena, such as X-ray data \cite{IronLine, Xray, Abdikamalov, deltaKerr} and gravitational waves. Establishing a framework for computing Quasi-Normal Modes (QNMs) for a KRZ metric could significantly expand the range of metrics suitable for QNM studies \cite{QNMKerr, Allahyari}.

Ultimately, the main motivation of this project is to approximate actual numerical metrics with an analytical form. Hence, conducting experiments on spacetimes with unknown exact expressions is essential. Without having expected results to compare errors against, it is possible to compare different orders of approximation instead, given the rapid and guaranteed convergence of the Konoplya-Rezzolla-Zhidenko parameterization. Computationally intensive projects like GRMHD simulations would be perfect testing grounds for the algorithm \cite{Hawley}.

\vspace{0.5cm}

{\bf Acknowledgments --}
We would also like to thank Alexander Zhidenko for useful comments and suggestions.
This work was supported by the Natural Science Foundation of Shanghai, Grant No.~22ZR1403400, and the National Natural Science Foundation of China (NSFC), Grant No.~12250610185 and 12261131497.
O.M. acknowledges also the support from the Shanghai Government Scholarship (SGS).
T.M. acknowledges also the support from the China Scholarship Council (CSC), Grant No.~2022GXZ005433.

\bibliography{ref.bib}

\end{document}